\documentclass[aps,pre,reprint,groupedaddress,showpacs]{revtex4-1}
\usepackage[english]{babel}
\usepackage[latin2]{inputenc}
\usepackage{graphicx}
\usepackage{amsmath}
\usepackage{verbatim}

\begin{document}

\title{A model for effective interactions in binary colloidal systems of soft particles}
\author{M. Majka}
\email{maciej.majka@uj.edu.pl}
\affiliation{Marian Smoluchowski Institute of Physics, Jagiellonian University, Reymonta 4, 30-059 Kraków, Poland}
\author{P. F. G\'{o}ra}
\affiliation{Marian Smoluchowski Institute of Physics, Jagiellonian University, Reymonta 4, 30-059 Kraków, Poland}


\begin{abstract}
While the density functional theory with integral equations techniques are very efficient tools in numerical analysis of complex fluids, an analytical insight into the phenomenon of effective interactions is still limited. In this paper we propose a theory of binary systems which results in a relatively simple analytical expression combining arbitrary microscopic potentials into the effective interaction. The derivation is based on translating many particle Hamiltonian including particle-depletant and depletant-depletant interactions into the occupation field language. Such transformation turns the partition function into multiple Gaussian integrals, regardless of what microscopic potentials are chosen. In result, we calculate the effective Hamiltonian and discuss when our formula is a dominant contribution to the effective interactions. Our theory allows us to analytically reproduce several important characteristics of systems under scrutiny. In particular, we analyze the effective attraction as a demixing factor in the binary systems of Gaussian particles, effective interactions in the binary mixtures of Yukawa particles and the system of particles consisting of both repulsive core and attractive/repulsive Yukawa interaction tail, for which we reproduce the 'attraction-through-repulsion' and  'repulsion-through-attraction' effects.
\end{abstract}


\maketitle

\section{Introduction} \label{sec:introduction}

Effective interactions are of central interest for the soft matter physics \cite{bib:likos}, especially in colloid studies. Their significance is enormous, since they are essential for spontaneous self-organization, they play a key-role in the polymer studies \cite{bib:likos} as well as in gel- and glass- forming research \cite{bib:nature,bib:weeks}. They are also important for molecular biophysics \cite{bib:marenduzzo} and find multiple applications in nano-technology \cite{bib:softnano}. Qualitatively similar phenomena of size separation are also encountered in the vibrated granular matter research \cite{bib:kudrolli, bib:aumaitre}.

A comprehensive introduction to the topic of effective interactions can be found in \cite{bib:likos,bib:hansen,bib:lekkerkerker}. The first attempt to describe the effective interactions in colloids was accomplished in late '50 by Asakura and Oosawa \cite{bib:asakura1,bib:asakura2} and, separately, by Vrij \cite{bib:vrij}. They considered a system of small and big hard spheres and identified 'excluded volume interaction' arising from entropy gain for the smaller particles (depletant) when the bigger particles are clustered. Their classical approach is continued until today, especially in the context of non-spherical particles (e.g. \cite{bib:yaman,bib:lang}). At the advent of the optical tweezers technology \cite{bib:grier}, the depletion interactions became accessible for direct measurements. The predictions of Asakura-Oosawa model have been confirmed, especially at low volume fraction packing for hard-spheres solution and in the semi-dilute regime of hard-sphere in polymers solutions \cite{bib:yodh}. However, it is known that for densely packed systems or when non-hard-sphere interactions are present, the Asakura-Oosawa theory becomes insufficient \cite{bib:likos}.

One reason is that at high volume fraction packing, the system approaches glassy transition in which mobility is reduced and strong translational spatial correlations appear. This has been observed both experimentally \cite{bib:weeks,bib:donth,bib:impendspec} and via simulations \cite{bib:donati,bib:doliwa,bib:mitus}. On the other hand, systems with non-trivial interactions can be constructed. This includes charged particles which interact over a range of a few diameters \cite{bib:crocker}, polymer coated particles interacting via mushroom-like potentials \cite{bib:DNA} or polymer coils, which in a good solvent behave effectively like soft, Gaussian-profiled particles \cite{bib:gaussians}. The systematic molecular dynamics simulations for various combinations of repulsions and attractions in the systems have also shown that unexpected effects can be encountered, e.g. effective repulsion arising from attractive microscopic potentials or the effective attraction induced by repulsive microscopic potentials \cite{bib:louise}.
 
A general theory capable of handling these phenomena has been proposed by Dijkstra in '90 \cite{bib:dijkstra1}. In this approach, a partition function for the system with arbitrary chosen particle-depletant and depletant-depletant interaction is systematically expanded in terms of Meyer bond functions, related to 0-body, two-body, three-body etc. interactions \cite{bib:likos, bib:hansen}. While Meyer bond expansion is, in principle, exact, including high-order terms is usually challenging or even intractable, due to their mathematical form and non-perturbative character. Therefore, a class of approximated techniques based on integral equations, closure relations and utilizing various density correlation functions has been also proposed \cite{bib:likos}. They  became a standard tool in the field, especially efficient in the numerical analysis of various systems e.g. \cite{bib:gaussians, bib:dijkstra2, bib:heinonen}. Nevertheless, an analytical form of effective interactions is known only for several model systems (see \cite{bib:lekkerkerker} for review) and similar results for complex fluids are rather scarce (e.g. \cite{bib:yatsenko,bib:mccarty}). 

While it is notoriously challenging to predict the effective interactions from arbitrary microscopic potentials, a simplified, tough analytical theory could find multiple applications in the colloid research, e.g. in the high-level solution design or in the context of Langevin dynamics simulation (e.g.\cite{bib:sagues,bib:majka1,bib:majka2,bib:majka3}). In this paper, we propose such a new theory which offers both generality and comprehensible analytical form. 

\subsection{Effective interactions in occupation functional theory}
We consider a binary system of spherically symmetric particles with arbitrarily chosen microscopic potentials. In our approach, we introduce the so-called occupation functional (representing a number of particles at every position) and translate the semi-grand canonical ensemble into the path-integral problem related to this functional. Regardless of microscopic potentials, this method turns the partition function into multiple Gaussian integrals. There are two major advantages of this transformation. On the one hand, we are able to identify and factorize a closed-form formula contributing to effective interactions, which is exact.  On the other hand, we can efficiently approximate the effective Hamiltonian in order to identify further contributions and provide the criteria under which the exact part is dominant.

In our model, similarly to \cite{bib:likos} and \cite{bib:dijkstra1}, we consider two distinct species of particles in the $D$-dimensional volume $\Omega=L^D$. The system has temperature $T$ and we will denote $\beta=(k_B T)^{-1}$, where $k_B$ is the Boltzmann constant. We will also use $h$ to denote the Planck constant. In the system, there are $N_1$ particles of the first kind and we denote the position and momentum of  $i$-th particle with $\textbf{R}_i$  and $\textbf{P}_i$, respectively. The microscopic potential between these particles reads $U_{RR}(|\textbf{R}_i-\textbf{R}_j|)$ and  effective interaction will be derived for this species. The second species, identified as depletant, consists of $N_2$ particles, which interact via potential $V(|\textbf{r}_i-\textbf{r}_j|)$ and their positions and momenta are denoted with $\textbf{r}_i$ and $\textbf{p}_i$. We will use the grand canonical ensemble for depletant, so we associate a chemical potential $\mu$ with this species. Both types of particles cross-interact via potential $U(|\textbf{R}_i-\textbf{r}_j|)$. The masses of colloid and depletant particles are $M$ and $m$, respectively. The total Hamiltonian of the system in its initial form has three contributions:
\begin{equation}
H_{tot}=H_{RR}+H_{rR}+H_{rr} \label{eq:H_tot}
\end{equation}
where:
\begin{align}
H_{RR}&=\sum_{i}^{N_1}\frac{\textbf{P}_i^2}{2M}+\frac{1}{2}\sum_{\substack{i,j\\i\neq j}}^{N_1}U_{RR}(|\textbf{R}_i-\textbf{R}_j|) \\
H_{rR}&=\sum_{i}^{N_1}\sum_{j}^{N_2} U(|\textbf{R}_i-\textbf{r}_j|)\\
H_{rr}&=\sum_{i}^{N_2}\frac{\textbf{p}_i^2}{2m}+\frac{1}{2}\sum_{\substack{i,j\\i\neq j}}^{N_2}V(|\textbf{r}_i-\textbf{r}_j|) \label{eq:H_rr}
\end{align}
Let us introduce a pair of Fourier Transforms:
\begin{gather*}
\mathcal{U}(k)=\int_{\Omega} d\textbf{r} e^{\imath \bf{k}\bf{r}}U(r)\\
\mathcal{V}(k)=\int_{\Omega} d\textbf{r} e^{\imath \bf{k}\bf{r}}V(r)
\end{gather*}
We will show that the effective interaction between a pair of colloid particles positioned at $\textbf{R}_i$ and $\textbf{R}_j$ can be expressed by:
\begin{equation}
U_{eff}(\textbf{R}_i-\textbf{R}_j)=-\frac{1}{(2\pi)^D}\int_{\tilde \Omega}d\textbf{k} e^{\imath \bf{k}(\textbf{R}_i-\textbf{R}_j)}\frac{|\mathcal{U}(k)|^2}{\mathcal{V}(k)} \label{eq:Ueff_int}
\end{equation}
This result is exact and sufficient to reproduce many important characteristics of binary mixtures. By calculating the approximated form of total effective Hamiltonian we will also show that there are other sources of effective interactions and we will provide a general criterion under which $U_{eff}(\textbf{R}_i-\textbf{R}_j)$ is dominant.

The paper is organized as follows: in Sections \ref{sec:partition_fun}-\ref{sec:non_negative} we introduce our framework of occupation functional, in Section \ref{sec:effective_int} the formula for $U_{eff}(\textbf{R}_i-\textbf{R}_j)$ is derived, in Section \ref{sec:approximations} the approximated partition functions is calculated and Sections \ref{sec:tot_hamiltonian} concludes on the effective Hamiltonian and the accuracy of our model. The assumptions and caveats regarding derivation are summarized in Section \ref{sec:caveats}. Section \ref{sec:applications} contains the examples of application for our theory. This includes the binary mixtures of Gaussian particles (\ref{sec:gaussian}), mixtures of Yukawa particles (\ref{sec:yukawa}) and Yukawa particles with impenetrable cores (\ref{sec:yukawa_hs}) for which the effects of 'attraction-through-repulsion' and 'repulsion-through-attraction' are reproduced. 

\section{Model derivation}\label{sec:model_derivation}

\subsection{System partition function} \label{sec:partition_fun}

In order to begin the derivation of our model we have to specify the partition function of the system. Our aim is to apply a new way to integrate-out the depletant degrees of freedom. In result, the effective Hamiltonian will be derived from the remaining expression. The initial Hamiltonian $H_{tot}$ is defined by equations \eqref{eq:H_tot} to \eqref{eq:H_rr}. It is feasible to rewrite $H_{rr}$ in the following manner:
\begin{equation}
H_{rr}=\sum_{i}^{N_2}\frac{\textbf{p}_i^2}{2m}+\frac{1}{2}\sum_{i,j}^{N_2}V(|\textbf{r}_i-\textbf{r}_j|)-\frac{N_2}{2}V(0)
\end{equation}
which explicitly introduces $V(0)$.

For Hamiltonian $H_{tot}$ we introduce a mixed ensemble $\Xi_{tot}$, which is the grand canonical ensemble for depletant and the canonical ensemble for colloid particles. Written in standard space-momentum coordinates $\{\textbf{P}_i,\textbf{R}_i\}_{N_1}$ and $\{\textbf{p}_i,\textbf{ r}_i\}_{N_2}$ the mixed ensemble reads:
\begin{equation}
\Xi_{tot}=\prod_i^{N_1} \int d\textbf{P}_i d\textbf{R}_i \frac{\exp\left( -\beta(H_{RR}-\frac{1}{\beta}\ln\Xi)\right)}{N_1!h^{DN_1}}
\end{equation}
where:
\begin{equation}
\Xi=\sum_{N_2=0}^{+\infty} \int d\textbf{p}_i d\textbf{r}_i \frac{\exp\left(-\beta (H_{rR}+H_{rr}-\mu N_2)\right)}{N_2!h^{DN_2}} \label{eq:Xi}
\end{equation}
According to \cite{bib:likos}, the term:
\begin{equation}
U_{eff}^{tot}=-\frac{1}{\beta}\ln\Xi
\end{equation}
act as an additional potential for colloid particles and this is the source of effective interactions. Therefore, calculating $\Xi$ is of central interest for us.

Another step is to integrate out the momenta $\textbf{p}_j$ in $\Xi$:
\begin{equation}
\prod_j^{N_2}\int_{-\infty}^{+\infty}d\textbf{p}_j e^{-\frac{\beta}{2m}\textbf{p}_j^2}=\left(\frac{2\pi m}{\beta}\right)^{\frac{D}{2}N_2}
\end{equation}
Now, it is possible to rearrange $\Xi$ into the form:
\begin{equation}
\Xi=\sum_{N_2}\frac{1}{L^{DN_2}}\prod_{j}^{N_2}\int d\textbf{r}_j\frac{\exp\left(-\beta(\mathcal{H}-\tilde \mu N_2\right))}{\Gamma(N_2+1)} \label{eq:Xi_ini}
\end{equation}
where $\Gamma(\dots)$ is the Euler Gamma function replacing the factorial and:
\begin{gather}
\mathcal{H}=\frac{1}{2}\sum_{i,j}^{N_2}V(|\textbf{r}_i-\textbf{r}_j|)+\sum_{i}^{N_1}\sum_{j}^{N_2} U(|\textbf{R}_i-\textbf{r}_j|)\\
\tilde \mu=\mu+\frac{1}{2}V(0)+\frac{D}{2\beta}\ln\frac{2\pi L^2 m}{\beta h^2}
\end{gather}
The partition function $\Xi$ given in the form \eqref{eq:Xi_ini} is ready to be translated into the occupation field representation.

\subsection{Occupation field representation}\label{sec:occupation_field}
Let us consider a scalar field which assigns the number of depletant particles $\alpha(\textbf{r})$ at certain position $\textbf{r}$ to this position. The number of depletant particles reads:
\begin{equation}
N_2=\int_{\Omega} d\textbf{r} \alpha(\textbf{r}) \label{eq:N2}
\end{equation}
If $\alpha(\textbf{r})$ particles occupies a position $\textbf{r}$ and $\alpha(\textbf{r}')$ occupies a position $\textbf{r}'$, then the energy of interaction between the sites $\textbf{r}$ and $\textbf{r}'$ is equal to $\alpha(\textbf{r})\alpha(\textbf{r}')V(|\textbf{r}-\textbf{r}'|)$. Therefore, we can use $\alpha(\textbf{r})$ to translate interaction terms in the following manner:
\begin{gather}
\sum_{i,j}^{N_2}V(|\textbf{r}_i-\textbf{r}_j|)=\iint_{\Omega} d\textbf{r} d\textbf{r}' \alpha(\textbf{r})\alpha(\textbf{r}') V(|\textbf{r}-\textbf{r}'|) \\
\sum_{i}^{N_1}\sum_{j}^{N_2} U(|\textbf{R}_i-\textbf{r}_j|)=\sum_{i}^{N_1}\int_{\Omega} d\textbf{r}\alpha(\textbf{r}) U(|\textbf{R}_i-\textbf{r}|) \label{eq:Utranslated}
\end{gather}
In principle, $\alpha(\textbf{r})$ takes only discrete values $0,1,2,\dots$, but we will allow it to vary continuously. 

The formulas \eqref{eq:N2}-\eqref{eq:Utranslated} suggest that we can understand $\mathcal{H}$ and $N_2$ as the functionals of $\alpha(\textbf{r})$. In turn, we could replace the multiple integrations in \eqref{eq:Xi_ini} with a functional integral with respect to $\alpha(\textbf{r})$, namely:
\begin{equation}
\Xi \to \int \mathcal{D}[\alpha]\frac{\exp\left(-\beta(\mathcal{H}+\tilde \mu N_2) \right)}{\Gamma(N_2+1)}
\end{equation}

The path integral can be specified as the integral with respect to Fourier series coefficients of $\alpha(\textbf{r})$ \cite{bib:pathint}:
\begin{equation}
\alpha(\textbf{r})=\frac{1}{\Omega}\sum_{\textbf{n}\in Z^D}a_n e^{\imath \frac{2\pi}{L}\bf{nr}} \label{eq:fourier}
\end{equation}
Here $\textbf{n}$ is a $D$-dimensional vector, which components varies discretely from $-\infty$ to $+\infty$. Therefore, we shall denote the set of index vectors $\textbf{n}$ with $Z^D$. The Fourier series expansion of $\alpha(\textbf{r})$ requires us to assume periodic boundary conditions. Since the field $\alpha(\textbf{r})$ is real, the symmetry $a_{-\textbf{n}}=a_{\textbf{n}}^*$ is also required. The $a_0$ coefficient has a specific interpretation:
\begin{equation}
a_0=\int_{\Omega}d\textbf{r}\alpha(\textbf{r})=N_2
\end{equation}
Additionally, we have to assume that potentials $U(\textbf{r})$ and $V(\textbf{r})$ are also periodic over length $L$, which should be of little influence if the range of those potentials is much shorter than $L$. If so, then Fourier series expansion \eqref{eq:fourier} simplifies the interaction terms:
\begin{gather}
\iint_{\Omega} d\textbf{r} d\textbf{r}' \alpha(\textbf{r})\alpha(\textbf{r}') V(|\textbf{r}-\textbf{r}'|)=\frac{1}{\Omega}\sum_{\textbf{n}\in Z^D}|a_\textbf{n}|^2\mathcal{V}_{\textbf{n}}\\
\sum_{i}^{N_1}\int_{\Omega} d\textbf{r}\alpha(\textbf{r}) U(|\textbf{R}_i-\textbf{r}|)=\frac{1}{\Omega}\sum_{\textbf{n}\in Z^D}a_{\textbf{n}}\sum_i\mathcal{U}_{\textbf{n}}^{(i)}
\end{gather}
where:
\begin{gather}
\mathcal{V}_{\textbf{n}}=\int_{\Omega}d\textbf{r}e^{\imath \frac{2\pi}{L}\textbf{nr}}V(r) \\
\mathcal{U}_{\textbf{n}}^{(i)}=\int_{\Omega}d\textbf{r}e^{\imath \frac{2\pi}{L}\textbf{nr}}U(|\textbf{R}_i-\textbf{r}|)
\end{gather}
From these formulas it follows that:
\begin{equation}
\begin{split}
&\mathcal{H}-\tilde \mu N_2=\\
&=\frac{1}{2\Omega}\sum_{\textbf{n}\in Z^D}|a_{\textbf{n}}|^2\mathcal{V}_{\textbf{n}}+\frac{1}{\Omega}\sum_{\textbf{n}\in Z^D}a_{\textbf{n}}\left( \sum_{i}\mathcal{U}_{\textbf{n}}^{(i)}-\tilde\mu \Omega\delta_{\textbf{n},0}\right)
\end{split}
\end{equation}
which can be further rearranged into:
\begin{equation}
\begin{split}
\mathcal{H}-\tilde \mu N_2=&\sum_{\textbf{n}\in Z^D}\frac{\mathcal{V}_\textbf{n}}{2\Omega}\left|a_ {\textbf{n}}+\frac{\sum_{i} \mathcal{U}_{-\textbf{n}}^{(i)}-\tilde\mu \Omega\delta_{\textbf{n},0}}{\mathcal{V}_\textbf{n}}\right|^2 \\
&- \sum_{\textbf{n}\in Z^D}\frac{\left|\sum_{i}\mathcal{U}_{\textbf{n}}^{(i)}-\tilde\mu \Omega\delta_{\textbf{n},0}\right|^2}{2\Omega\mathcal{V}_\textbf{n}} \label{eq:Hfactorized}
\end{split}
\end{equation}
and finally the path integral is specified as:
\begin{equation}
\Xi=\prod_{\textbf{n}\in Z^D}\int da_{\textbf{n}}\frac{\exp\left(-\beta(\mathcal{H}+\tilde \mu N_2) \right)}{\Gamma(a_0+1)} \label{eq:Xi_a}
\end{equation}
In the above formula, we intentionally omit writing the limits of integration since they need to be discussed in grater detail in the following section.

\subsection{Nonnegative fields from Fourier modes}\label{sec:non_negative}
 In principle, the occupation field $\alpha(\textbf{r})$ should be non-negative. Unfortunately, a field constructed according to \eqref{eq:fourier} from the arbitrarily chosen values of $a_{\textbf{n}}$ does not necessarily meet this requirement. However, it is always true that $a_0\ge0$, since it is the number of depletant particles. Therefore, for any values of $a_{\textbf{n}\neq0}$ we can choose such $a_0$ that ensures $\alpha(\textbf{r})$  is non-negative. More precisely, we can write:
\begin{equation}
\tilde \alpha(\textbf{r})=\sum_{\textbf{n}\in Z^D\setminus 0}a_n e^{\imath \frac{2\pi}{L}\bf{nr}} \label{eq:nonneg1}
\end{equation}
where $\setminus0$ indicates the exclusion of $a_0$. $\tilde \alpha(\textbf{r})$ is a real function and, necessarily:
\begin{equation}
\int_{\Omega}d\textbf{r}\tilde \alpha(\textbf{r})=0 \label{eq:nonneg_prop}
\end{equation}
This property means that $\tilde \alpha(\textbf{r})$ has to take both negative and nonnegative values for different $\textbf{r}$, so integral \eqref{eq:nonneg_prop} is 0. Therefore, there must exist a global minimum of $\tilde \alpha(\textbf{r})$ and $\tilde \alpha(\textbf{r})$ is negative in this minimum. Finally, for:
\begin{equation}
a_0\ge m=-\textrm{min}_{\textbf{r}}(\tilde\alpha(\textbf{r})) \label{eq:nonneg2}
\end{equation}
the occupation field $\alpha(\textbf{r})$ is non-negative. Here, we denote  the global minimum of $\tilde \alpha(\textbf{r})$ with respect to $\textbf{r}$ by $\textrm{min}_{\textbf{r}}\left(\tilde \alpha(\textbf{r}) \right)$. The limit $m$ can be also rewritten in the following form:
\begin{equation}
m=-\sum_{\textbf{n}\in Z^D\setminus 0}a_{\textbf{n}}e^{\imath \frac{2\pi}{L}\textbf{nr}(a_{\textbf{n}})} \label{eq:m1}
\end{equation}
where $\textbf{r}(a_{\textbf{n}})$ is the position of global minimum as a function of $a_{\textbf{n}}$. $\textbf{r}(a_{\textbf{n}})$ can be determined from the equation:
\begin{equation}
\nabla_{r} \sum_{\textbf{n}\in Z^D\setminus 0}a_{\textbf{n}}e^{\imath \frac{2\pi}{L}\textbf{nr}}=0
\end{equation}

Concluding this section, we can choose the limits of integration for $a_{\textbf{n}\neq 0}$ as $\pm\infty$ and the limits for $a_0$ as $[m,+\infty)$. However, $m$ is now a function of $a_{\textbf{n}\neq0}$, which fixes the order of integrals in \eqref{eq:Xi_a}. Let us combine \eqref{eq:Hfactorized} and \eqref{eq:Xi_a} to write $\Xi$ in the following form:
\begin{equation}
\Xi=e^{-\beta \Phi }\prod_{\textbf{n}\in Z^D\setminus 0}I_{\textbf{n}} I_0 \label{eq:Xi_I1}
\end{equation}
in which:
\begin{equation}
\Phi=-\sum_{\textbf{n}\in Z^D}\frac{\left|\sum_{i}\mathcal{U}_{\textbf{n}}^{(i)}-\tilde\mu \Omega\delta_{\textbf{n},0}\right|^2}{2\Omega\mathcal{V}_\textbf{n}} \label{eq:Phi}
\end{equation}
and:
\begin{gather}
I_{\textbf{n}\neq0}=\int_{-\infty}^{+\infty}da_{\textbf{n}}\exp\left(-\frac{\beta \mathcal{V}_{\textbf{n}}}{2\Omega}\left|a_{\textbf{n}}+\frac{\sum_{i}\mathcal{U}_{-\textbf{n}}^{(i)}}{\mathcal{V}_\textbf{n}}\right|^2 \right) \label{eq:In} \\
I_0(m)=\int_{m}^{+\infty}da_0\frac{\exp\left(-\frac{\beta \mathcal{V}_0}{2\Omega}\left|a_0+\frac{\sum_{i}\mathcal{U}_{0}^{(i)}-\tilde\mu \Omega}{\mathcal{V}_0} \right|^2 \right) }{\Gamma(a_0+1)} \label{eq:I0}
\end{gather}
For the sake of more compact notation, we will denote: 
\begin{align}
&c_\textbf{n}=\frac{\sum_{i}\mathcal{U}_{-\textbf{n}}^{(i)}-\tilde\mu \Omega \delta_{0\textbf{n}}}{\mathcal{V}_\textbf{n}}& \gamma_\textbf{n}=\frac{\beta\mathcal{V}_\textbf{n}}{2\Omega} \label{eq:c_n}
\end{align}

\subsection{The effective interaction}\label{sec:effective_int}
In this section we will identify the exact part of effective interactions. We substitute now \eqref{eq:Xi_I1} into the formula for $U_{eff}^{tot}$, namely:
\begin{equation}
U_{eff}^{tot}=-\frac{1}{\beta}\ln \Xi=\Phi-\frac{1}{\beta}\ln\left(\prod_{\textbf{n}\in  Z^D\setminus 0} I_\textbf{n}I_0(m) \right)
\end{equation}
We will show that $\Phi$ gives rise to the effective interaction $U_{eff}(|\textbf{R}_i-\textbf{R}_j|)$. Expanding \eqref{eq:Phi} and taking the advantage of Kronecker delta, we arrive at:
\begin{equation}
\begin{split}
\Phi=&-\sum_{i\neq j}\sum_{\textbf{n}\in Z^D}\frac{\mathcal{U}_{\textbf{n}}^{(i)}\mathcal{U}_{-\textbf{n}}^{(j)}}{2\Omega\mathcal{V}_{\textbf{n}}}-\sum_{i}\sum_{\textbf{n}\in Z^D}\frac{|\mathcal{U}_{\textbf{n}}^{(i)}|^2}{2\Omega\mathcal{V}_{\textbf{n}}}+\\
&+\frac{2\tilde\mu\sum_{i}\mathcal{U}_0^{(i)}-\tilde \mu^2\Omega}{2\mathcal{V}_0}
\end{split}
\end{equation}
In order to process the three terms in $\Phi$, one can notice that:
\begin{equation}
\begin{split}
\mathcal{U}_{\textbf{n}}^{(i)}&=\int_{\Omega}d\textbf{r}e^{\imath \frac{2\pi}{L}\textbf{nr}}U(|\textbf{R}_i-\textbf{r}|) \\
&=e^{\imath\frac{2\pi}{L}\textbf{nR}_i}\int_{\Omega_i}d\textbf{r}e^{\imath\frac{2\pi}{L}\textbf{nr}}U(r)\\
&=e^{\imath\frac{2\pi}{L}\textbf{nR}_i}\mathcal{U}\left(\frac{2\pi}{L}\textbf{n}\right)
\end{split}
\end{equation}
Here $\Omega_i$ is a volume shifted by $\textbf{R}_i$. In the continuous limit of huge volume $L\to+\infty$ we can substitute $\textbf{k}=\frac{2\pi}{L}\textbf{n}$, so:
\begin{equation}
\mathcal{U}_{\textbf{n}}^{(i)}\to e^{\imath \textbf{kR}_i}\mathcal{U}(\textbf{k}) \label{eq:U_cont}
\end{equation}
Further, $\sum_\textbf{n}\to\frac{\Omega}{(2\pi)^D}\int_{\tilde \Omega}$ and $\Omega_i\to\Omega$, so $\mathcal{U}(\textbf{k})$ becomes a Fourier transform of $U(r)$. Similar considerations allow us to transform $\mathcal{V}_{\textbf{n}}$ into $\mathcal{V}(\textbf{k})$. Finally, in the continuous limit:
\begin{equation}
\begin{split}
-\sum_{\textbf{n}\in Z^D}\frac{\mathcal{U}_{\textbf{n}}^{(i)}\mathcal{U}_{-\textbf{n}}^{(j)}}{\Omega\mathcal{V}_{\textbf{n}}}&\to-\frac{1}{(2\pi)^D}\int_{\tilde \Omega}e^{\imath\textbf{k}(\textbf{R}_i-\textbf{R}_j)}\frac{|\mathcal{U}(\textbf{k})|^2}{\mathcal{V}(\textbf{k})}\\
&=U_{eff}(\textbf{R}_i-\textbf{R}_i)
\end{split} \label{eq:Ueff}
\end{equation}
Formula \eqref{eq:Ueff} constitutes the main result of this paper which is the expression for the effective interaction between two particles. Having established this result, it follows that:
\begin{equation}
-\sum_{i}\sum_{\textbf{n}\in Z^D}\frac{|\mathcal{U}_{\textbf{n}}^{(i)}|^2}{\Omega\mathcal{V}_{\textbf{n}}}\to\sum_i U_{eff}(0)=N_1 U_{eff}(0)
\end{equation}
and:
\begin{equation}
\frac{2\tilde\mu\sum_{i}\mathcal{U}_0^{(i)}-\tilde \mu^2\Omega}{2\mathcal{V}_0}\to\frac{2\tilde \mu N_1 \mathcal{U}(0)-\Omega\tilde \mu^2}{2\mathcal{V}(0)}
\end{equation}
In summary, we conclude that the general form of $\Phi$ reads:
\begin{equation}
\Phi=\frac{1}{2}\sum_{i\neq j}^{N_1}U_{eff}(\textbf{R}_i-\textbf{R}_j)+\frac{N_1}{2}U_{eff}(0)+\frac{2\tilde \mu N_1 \mathcal{U}(0)-\Omega\tilde \mu^2}{2\mathcal{V}(0)} \label{eq:Phi_fin}
\end{equation} 
Immediately one can recognize that we have obtained the effective interaction between every pair of particles, which is expected for the multi-particle system. This result is exact up to the approximations required to introduce the occupation number functional.

\subsection{Approximated calculation of  $\ln \prod I_\textbf{n}I_0(m)$}\label{sec:approximations}
Having found $\Phi$, we would also like to calculate the $\prod I_\textbf{n}I_0(m)$ to obtain total effective Hamiltonian. However, this can be completed only via certain approximations.

First of all, let us remind that, according to \eqref{eq:I0}, $I_0(m)$ reads:
\begin{equation*}
I_0(m)=\int_{m}^{+\infty}da_0\frac{e^{-\gamma_0(a_0+c_0)^2}}{\Gamma(a_0+1)}
\end{equation*}
In principle, $m$ is nonnegative and for such argument $I_0(m)$ is a decreasing function, reaching asymptotically 0 in the limit of $m\to+\infty$. $I_0(m)$ can have a well-defined kink at $m=-c_0$, provided that $c_0<0$ and $\gamma_0\gg1$. We will approximate now $I_0(m)$ up to the first order in the logarithmic derivative, namely:
\begin{equation}
I_0( m)=\exp\left(\ln I_0(0)+\frac{I'_0(0)}{I_0(0)} m+\dots \right)\simeq I_0(0)e^{\frac{I'_0(0)}{I_0(0)} m} \label{eq:expansion}
\end{equation}
This is accurate, provided that there is no kink for $m\in[0,+\infty)$, which requires that $c_0>0$. For more compact notation we denote:
\begin{equation}
\mathcal{I}_0=\frac{I'_0(0)}{I_0(0)}
\end{equation}

Under approximation \eqref{eq:expansion} and using expansion \eqref{eq:m1} for $m$, we can write:
\begin{equation}
\begin{split}
&\prod_{\textbf{n}\in Z^D\setminus 0}I_{\textbf{n}}I_0\approx\prod_{\textbf{n}\in Z^D\setminus 0} \int_{-\infty}^{+\infty}da_{\textbf{n}}I_0(0)\times \\
&\times \exp\left(-\gamma_{\textbf{n}} \left|a_{\textbf{n}}+c_{\textbf{n}} \right|^2-\mathcal{I}_0 a_{\textbf{n}}e^{\imath \frac{2\pi}{L}\textbf{nr}(a_{\textbf{n}})} \right)
\end{split} \label{eq:InI0_1}
\end{equation}
This expression is still dependent on $\textbf{r}(a_{\textbf{n}})$, which is an implicit function of $a_{\textbf{n}}$. To proceed, we will approximate $\textbf{r}(a_{\textbf{n}})$ by a constant value. One can notice that the quadratic term in \eqref{eq:InI0_1} is has the extreme value for $a_{\textbf{n}}=-c_{\textbf{n}}$ and we expect that the integral \eqref{eq:InI0_1} is dominated by the contribution from $a_{\textbf{n}}\approx-c_{\textbf{n}}$. Let us transform the integration variables:
\begin{equation}
\Delta a_{\textbf{n}}= a_{\textbf{n}}+c_{\textbf{n}}
\end{equation}
and approximate $m$ in the vicinity of $c_\textbf{n}$ up to first order in $\Delta a_{\textbf{n}}$:
\begin{equation}
-\sum_{\textbf{n}\in Z^D\setminus 0}a_{\textbf{n}}e^{\imath \frac{2\pi}{L}\textbf{nr}(a_{\textbf{n}})}\simeq \sum_{\textbf{n}\in Z^D\setminus 0}(c_{\textbf{n}}-\Delta a_{\textbf{n}})e^{\imath \frac{2\pi}{L}\textbf{nr}(c_{\textbf{n}})}
\end{equation}
Now, \eqref{eq:InI0_1} turns into:
\begin{equation}
\begin{split}
&\prod_{\textbf{n}\in Z^D\setminus 0}I_{\textbf{n}}I_0\approx \prod_{\textbf{n}\in Z^D\setminus 0} \exp \left( \mathcal{I}_0 c_\textbf{n} e^{\imath \frac{2\pi}{L}\textbf{nr}(c_{\textbf{n}})}\right) I_0(0)\times \\
&\times \int_{-\infty}^{+\infty}d\Delta a_{\textbf{n}} \exp\left(-\gamma_{\textbf{n}} \left|\Delta a_{\textbf{n}}\right|^2-\mathcal{I}_0\Delta a_{\textbf{n}}e^{\imath \frac{2\pi}{L}\textbf{nr}(c_{\textbf{n}})} \right)
\end{split} \label{eq:InI0_2}
\end{equation}
We can rearrange the quadratic expression in the exponent of \eqref{eq:InI0_2}:
\begin{equation}
\begin{split}
&\sum_{\textbf{n}\in Z^D\setminus 0}\gamma_{\textbf{n}} \left|\Delta a_{\textbf{n}}\right|^2+\mathcal{I}_0\sum_{\textbf{n}\in Z^D\setminus 0}\Delta a_{\textbf{n}}e^{\imath \frac{2\pi}{L}\textbf{nr}(c_{\textbf{n}})}=\\
&\sum_{\textbf{n}\in Z^D\setminus 0}\gamma_{\textbf{n}} \left|\Delta a_{\textbf{n}}+\mathcal{I}_0\frac{e^{-\imath \frac{2\pi}{L}\textbf{nr}(c_{\textbf{n}})}}{2\gamma_{\textbf{n}}}\right|^2-\sum_{\textbf{n}\in Z^D\setminus 0}\frac{\mathcal{I}_0^2}{4\gamma_{\textbf{n}}}
\end{split}\label{eq:quadratic}
\end{equation}
Finally, since the integration variable $\Delta a_{\textbf{n}}$ is complex, we introduce its polar parametrization:
\begin{equation}
\rho_{\textbf{n}}e^{\pm\imath\phi_{\textbf{n}}}=\Delta a_{\pm\textbf{n}}+\mathcal{I}_0\frac{e^{\mp \imath \frac{2\pi}{L}\textbf{nr}(c_{\textbf{n}})}}{2\gamma_{\pm \textbf{n}}} \label{eq:polar}
\end{equation}

Once \eqref{eq:quadratic} and \eqref{eq:polar} are applied to \eqref{eq:InI0_2}, the integrations can be performed, provided that all $\Re (\gamma_{\textbf{n}})>0$. The result reads:
\begin{equation}
\begin{split}
&\ln \prod_{\textbf{n}\in Z^D\setminus 0}I_{\textbf{n}}I_0\approx \ln I_0(0)+\sum_{\textbf{n}\in Z^D\setminus 0} \ln \frac{\pi}{\gamma_{\textbf{n}}}+ \\
&+\mathcal{I}_0\sum_{\textbf{n}\in Z^D\setminus 0} c_\textbf{n} e^{\imath \frac{2\pi}{L}\textbf{nr}(c_{\textbf{n}})}+\sum_{\textbf{n}\in Z^D\setminus 0}\frac{\mathcal{I}_0^2}{4\gamma_{\textbf{n}}} 
\end{split}\label{eq:InI0_3}
\end{equation}

\subsection{Total effective Hamiltonian and model accuracy}\label{sec:tot_hamiltonian}
Let us summarize the two preceding sections. The effective Hamiltonian of the entire system reads:
\begin{equation*}
H_{eff}=H_{RR}+\Phi-\frac{1}{\beta}\ln \prod_{\textbf{n}\in Z^D\setminus 0}I_{\textbf{n}}I_0
\end{equation*}
Turning \eqref{eq:InI0_3} into its continuous form we obtain the final expression for the effective Hamiltonian:
\begin{equation}
\begin{split}
&H_{eff}\approx H_{RR}+\Phi-\frac{1}{\beta}\left(\frac{\Omega}{(2\pi)^D}\int_{\tilde \Omega}d \textbf{k} \ln \frac{\pi}{\gamma(\textbf{k})}+ \right. \\
&+\frac{\mathcal{I}_0\Omega}{(2\pi)^D} \sum_i^{N_1}\int_{\tilde \Omega}d\textbf{k}e^{\imath \textbf{k}(\textbf{r}_{min}-\textbf{R}_i)}\frac{\mathcal{U}(\textbf{k})}{\mathcal{V}(\textbf{k})}-\frac{N_1\mathcal{I}_0\mathcal{U}(0)}{\mathcal{V}(0)}+\\
&\left.+\frac{\mathcal{I}_0^2\Omega}{4(2\pi)^D}\int_{\tilde \Omega}d \textbf{k}\frac{1}{\gamma(\textbf{k})}-\frac{\mathcal{I}_0^2}{4\gamma(0)}+\ln \frac{\gamma(0)I_0(0)}{\pi}\right)
\end{split} \label{eq:H_fin}
\end{equation}
where $r_{min}$ is the global minimum, found from the equation:
\begin{equation}
\nabla_{\textbf{r}}\sum_i^{N_1}\int_{\tilde \Omega}d\textbf{k}e^{\imath \textbf{k}(\textbf{r}-\textbf{R}_i)}\frac{\mathcal{U}(\textbf{k})}{\mathcal{V}(\textbf{k})}=0
\end{equation}
Further, according to \eqref{eq:Phi_fin}, the exact part of $H_{eff}$ reads:
\begin{equation*}
\Phi=\frac{1}{2}\sum_{i\neq j}^{N_1}U_{eff}(\textbf{R}_i-\textbf{R}_j)+\frac{N_1}{2}U_{eff}(0)+\frac{2\tilde \mu N_1 \mathcal{U}(0)-\Omega\tilde \mu^2}{2\mathcal{V}(0)}
\end{equation*}
where the effective interaction $U_{eff}(\textbf{R}_i-\textbf{R}_j)$ is defined by \eqref{eq:Ueff}.

Let us scrutinize the following term from $H_{eff}$:
\begin{equation}
\Delta U_{eff}=\frac{1}{\beta}\frac{\mathcal{I}_0\Omega}{(2\pi)^D} \sum_i^{N_1}\int_{\tilde \Omega}d\textbf{k}e^{\imath \textbf{k}(\textbf{r}_{min}-\textbf{R}_i)}\frac{\mathcal{U}(\textbf{k})}{\mathcal{V}(\textbf{k})} \label{eq:term}
\end{equation}
First of all, this term is an explicit function of $\textbf{R}_i$, which is in stark contrast to the Meyer bond expansion, in which such terms are excluded. This exclusion is motivated by the conservation of energy when the entire system is translated \cite{bib:likos}. However, in our case, the global translation: $\textbf{R}_i\to\textbf{R}_i+\delta$ yields $\textbf{r}_{min}\to \textbf{r}_{min}+\delta$, thus \eqref{eq:term} is, in fact, translationally invariant.

Secondly, one can notice that since $\textbf{r}_{min}$ is a function of $\textbf{R}_i$ itself, there is possibly an additional effective interaction embedded in $\Delta U_{eff}$. Therefore, $U_{eff}(\textbf{R}_i-\textbf{R}_j)$ is the dominant source of effective interactions provided that: 
\begin{equation}
U_{eff}(\textbf{R}_i-\textbf{R}_j)\gg\Delta U_{eff}
\end{equation}
Whether this relation is satisfied, it depends on both thermodynamical parameters and the choice of microscopic potentials, which makes it difficult to analyze in the general case. However, if this relation is seriously violated, one might attempt to estimate the influence of $\Delta U_{eff}$ on effective interactions from the following reasoning:
\begin{equation}
\begin{split}
|\Delta U_{eff}|&<\frac{1}{\beta}\frac{\mathcal{I}_0\Omega}{(2\pi)^D} \int_{\tilde \Omega}d\textbf{k}\sqrt{\left|\sum_i^{N_1}e^{\imath \textbf{k}(\textbf{r}_{min}-\textbf{R}_i)}\frac{\mathcal{U}(\textbf{k})}{\mathcal{V}(\textbf{k})} \right|^2}=\\
=&\frac{N_1^{1/2}}{\beta}\frac{\mathcal{I}_0\Omega}{(2\pi)^D} \int_{\tilde \Omega}d\textbf{k}\left| \frac{\mathcal{U}(\textbf{k})}{\mathcal{V}(\textbf{k})}\right| \sqrt{1+\sum_{i\neq j}^{N_1}\frac{e^{\imath \textbf{k}(\textbf{R}_j-\textbf{R}_i)}}{N_1}}\simeq \\
\simeq& \frac{1}{\beta}\frac{\mathcal{I}_0\Omega}{(2\pi)^D}\left( N_1^{1/2}\int_{\tilde \Omega}d\textbf{k}\left| \frac{\mathcal{U}(\textbf{k})}{\mathcal{V}(\textbf{k})}\right|+\right.\\
&\left.+\frac{1}{2N_1^{1/2}}\sum_{i\neq j}^{N_1}\int_{\tilde \Omega}d\textbf{k} e^{\imath \textbf{k}(\textbf{R}_j-\textbf{R}_i)}\left|\frac{\mathcal{U}(\textbf{k})}{\mathcal{V}(\textbf{k})}\right| \right)
\end{split}
\end{equation}
This formula also predicts the effective interactions, tough we expect it to be overestimated in this case.

\subsection{Caveats}\label{sec:caveats}
Throughout the derivation section we have introduced multiple assumptions which we would like to list and discuss now. The formula \eqref{eq:H_fin}, though seemingly very general, has numerous caveats.

First, we resort to the continuous representation of discrete expressions, tough we expect $\Omega$ to be finite. This is physically reasonable provided that the range of microscopic potentials is much smaller than the system size $L$. Another issue is that we require microscopic potentials $V(r)$ and $U(r)$ to posses their Fourier transforms. This rules out e.g. Lennard-Jones type potentials or polynomial potentials. Moreover, since \eqref{eq:Ueff} has a form of inverse Fourier transform, the integrand must be 'well-behaving' i.e. convergent for $k\to+\infty$ and without any essential singularities. Although certain mathematical tricks and approximations can be applied to circumvent such problems, this is the main reason why \eqref{eq:Ueff} is not a directly applicable 'silver bullet' formula.

The most startling concern is whether depletant-depletant potential can have a negative or partially negative Fourier Transform. Since each $I_{\textbf{n}\neq0}$ is the Gaussian integral, it would be divergent for $\mathcal{V}_{\textbf{n}}<0$ hence $\prod I_{\textbf{n}}I_0(m)\to+\infty$. In this case, $H_{tot}$ given by \eqref{eq:H_fin} is meaningless, but we will argue that $\Phi$ might still provide some useful information. In general, it is true that:
\begin{equation}
\ln I_{\textbf{n}}I_0(m)\le\ln I_{\textbf{n}}I_0(0)=\sum_{\textbf{n}\in Z^D\setminus 0}\ln I_{\textbf{n}}+\ln I_0(0) \label{eq:InI0_up}
\end{equation}
Now, let us consider an observable $O(\textbf{R}_i,\textbf{P}_i)$ and its average:
\begin{equation}
\bar O=\frac{\int d\textbf{P}_id\textbf{R}_i O(\textbf{R}_i,\textbf{P}_i)\exp(-\beta H_{tot})}{\int d\textbf{P}_id\textbf{R}_i \exp(-\beta H_{tot})}\label{eq:O_av}
\end{equation}
We can use \eqref{eq:InI0_up} to approximate $H_{tot}$, namely:
\begin{equation}
H_{tot}\approx H_{RR}+\Phi-\frac{1}{\beta}\sum_{\textbf{n}\in Z^D\setminus 0}\ln I_{\textbf{n}}-\frac{1}{\beta}\ln I_0(0) \label{eq:H_up}
\end{equation}
From \eqref{eq:In} it follows that $I_{\textbf{n}}$ is independent from $\textbf{R}_i$ for the properly shifted integration variable. Applying \eqref{eq:H_up} to \eqref{eq:O_av}, one can see that:
\begin{equation}
\bar O=\frac{\int d\textbf{P}_id\textbf{R}_i O(\textbf{R}_i,\textbf{P}_i)\exp \left(-\beta(H_{RR}+\Phi \right))}{\int d\textbf{P}_id\textbf{R}_i \exp \left(-\beta(H_{RR}+\Phi)\right)}
\end{equation} 
which is independent from divergent $I_{\textbf{n}}$. This reasoning, although not very rigorous, suggest that $\Phi$ and $U_{eff}(\textbf{R}_i-\textbf{R}_j)$ might be accurate predictions for potentials with partially negative $\mathcal{V}(\textbf{k})$ and can be useful in determining the mean values. 

\section{Applications}\label{sec:applications}
\subsection{Systems under scrutiny}
In this section, we apply $U_{eff}(\textbf{R}_i-\textbf{R}_j)$ given by \eqref{eq:Ueff} to analyze effective interactions in various systems. Three classes of systems will be discussed. We begin with idealized case of point-like depletant and identify the limit of no effective interactions in binary system. Further, we focus on the binary mixtures of Gaussian particles, predicting effective interactions and analyzing effective attraction as a driving force behind demixing. Finally, we examine systems of particles described with Yukawa potential and Yukawa interaction tail with repulsive core. In the latter case, we can qualitatively reproduce the effects of 'attraction-through-repulsion'/'repulsion-through-attraction' observed in such systems earlier, via simulations \cite{bib:louise}.

\subsection{Point-like particles}
The simplest, though idealized model of depletant are point-like particles. The relevant potential and its Fourier Transform read:
\begin{align}
V(r)=V_{core}\delta(\textbf{r})&&\mathcal{V}(k)=V_{core}
\end{align}
In this case, formula \eqref{eq:Ueff_int} reduces to:
\begin{equation}
U_{eff}(\Delta R)=-\frac{1}{(2\pi)^DV_{core}}\int_{\tilde \Omega} d\textbf{k} e^{\imath \bf{k \Delta R}}|\mathcal{U}(k)|^2 \label{eq:Upoint1}
\end{equation}
Since $|\mathcal{U}(k)|^2=\mathcal{U}(k)\mathcal{U}^*(k)$, the expression \eqref{eq:Upoint1} is the inverse Fourier Transform of a product of two Fourier Transforms, so, by the power of convolution theorem:
\begin{equation}
U_{eff}(\Delta R)=-\frac{1}{V_{core}}\int_{\Omega} d\textbf{r} U(|\textbf{r}|) U(|\textbf{r}+\bf{\Delta R}|) \label{eq:Upoint2}
\end{equation}
which is simply the auto-convolution of $U$. 

Having established \eqref{eq:Upoint2}, we can now imagine a system consisting of ideal hard-spheres (HS) of radius $\sigma$ and a point-like depletant. In this case, $U_{eff}(\Delta R)=0$ if $\Delta R>2\sigma$ and $U_{eff}(\Delta R)=+\infty$ for $\Delta R<2\sigma$. However, $\Delta R<2\sigma$ is physically inaccessible for HS, so the effective interaction in such system is necessarily 0. This conclusion is in agreement with qualitative reasoning: point-like particles take no entropic advantage from any configuration of impenetrable spheres and hence no effective interaction should occur. Therefore, point-like depletant and HS mixture is a limit in which no effective interaction occurs. 

The situation is qualitatively different for soft-core $U(r)$. In this case, the non-zero auto-convolution would vary continuously for different $\Delta R$, which means that even the point-like particles can induce effective interactions, provided that $U(r)$ is 'soft'. 

\subsection{Gaussian particles and demixing of binary mixtures}\label{sec:gaussian}
Particles interacting via the Gaussian potential are a typical example of soft particles and can be analyzed within our framework. We will take advantage of the fact that the Fourier Transform of Gaussian is also a Gaussian function: 
\begin{align}
G(r)=\epsilon e^{-\frac{1}{2}\frac{r^2}{\sigma^2}}&&\mathcal{G}(k)=\epsilon (2\pi)^{D/2}\sigma^D e^{-\frac{1}{2}k^2\sigma^2}
\end{align}
Gaussian potential has been identified as the accurate approximation of the interaction between two isolated polymers in a good solvent, both for identical \cite{bib:gaussians} and non-identical \cite{bib:gaussian2} chains. Therefore, Gaussian-core model is a well established coarse-grained description of polymer solutions \cite{bib:likos} both in homogeneous and non-homogeneous case \cite{bib:gaussian_fluid}. In particular, it has been found that the binary mixtures of Gaussian particles can undergo size separation transition \cite{bib:gaussian_fluid,bib:gauss_separation}, similarly to polymer blends. 

\begin{figure}
\includegraphics[width=0.95\linewidth]{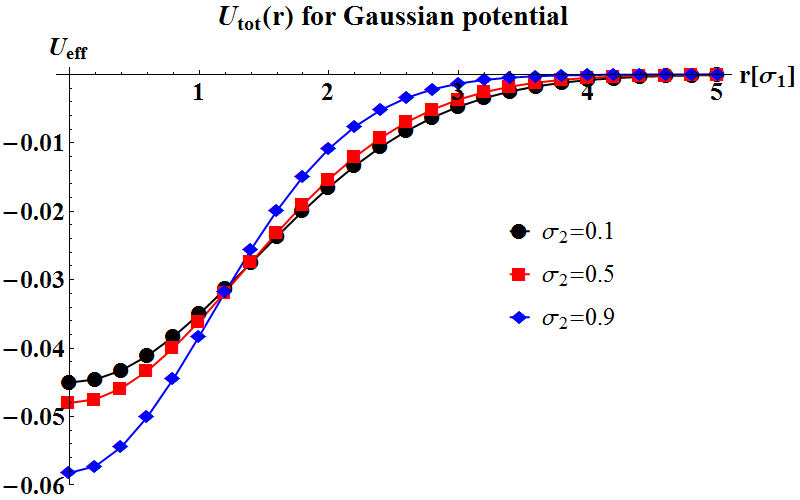}
\caption{Effective interaction between Gaussian particles, according to formula \eqref{eq:Ueff_gauss}. $\sigma_1$ is the unit length and the scaling reads $\epsilon_1^2/\epsilon_0=1$. $U_{eff}$ is a negative Gaussian function for every $\sigma_2$. \label{fig:gauss}}
\end{figure}

In our model we assume the binary mixture of different-sized Gaussian particles and assign index 1 to big-small interaction and 2 to small-small interaction. Then, the effective interaction, according to \eqref{eq:Ueff}, reads:
\begin{equation}
\begin{split}
U_{eff}(\Delta R)&=-\frac{1}{(2\pi)^D}\int_{\tilde \Omega} d\textbf{k} e^{\imath \bf{k \Delta R}}\frac{|\mathcal{G}_1(k)|^2}{\mathcal{G}_2(k)}\\
&=-\frac{\epsilon_1^2}{\epsilon_2}\frac{\sigma_1^{2D}}{(2\pi )^{D/2}\sigma_2^D}\frac{e^{-\Delta R^2/(4\sigma_1^2-2\sigma_2^2)}}{(2\sigma_1^2-\sigma_2^2)^{D/2}} \label{eq:Ueff_gauss}
\end{split}
\end{equation}
$U_{eff}(\Delta R)$ proves to be a renormalized Gaussian, but, since $\epsilon_i>0$, it is always negative. Examples of this interaction are presented in Fig. \ref{fig:gauss}.

\begin{figure}
\includegraphics[width=0.9\linewidth]{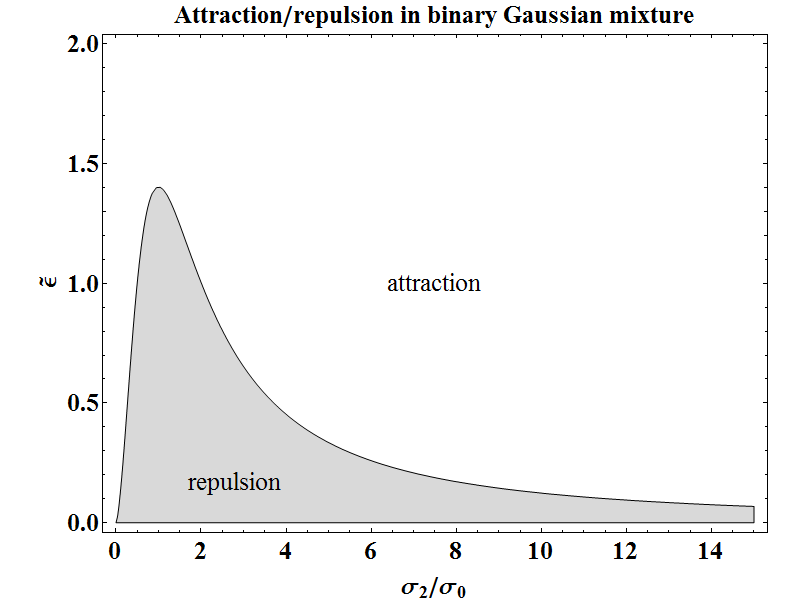}
\caption{Effective attraction in binary mixtures of Gaussian particles as a driving force behind phase separation. This plot visualizes inequality \eqref{eq:separation2} for $D=3$, where $\tilde \epsilon=\epsilon_1/\sqrt{\epsilon_0\epsilon_2}$ and $c=\sigma_2/\sigma_0$. Shaded region - total interaction is purely repulsive, no driving force for de-mixing. Plain region - total interaction has an attractive tail stimulating de-mixing.  \label{fig:separation}}
\end{figure}

Result \eqref{eq:Ueff_gauss} suggests that the total interaction between bigger particles (i.e. $U_{RR}(\Delta R)+U_{eff}(\Delta R)$) can include an attractive tail, provided that for certain choice of parameters there exist such $\Delta R$ that the effective interaction prevails over $U_{RR}(\Delta R)$. Possibly, such a tail could drive the separation process. Let the interaction between bigger particles read:
\begin{equation}
U_{RR}(\Delta R)=G_0(\Delta R)=\epsilon_0 e^{-\frac{1}{2}\frac{\Delta R^2}{\sigma_0^2}}
\end{equation}
The attractive tail will be present if the following inequality has a solution in $\Delta R$:
\begin{equation}
G_0(\Delta R)+U_{eff}(\Delta R)<0
\end{equation}
which can be reduced to:
\begin{equation}
\begin{split}
&\Delta R^2\left(\frac{1}{4\sigma_1^2-2\sigma_2^2}-\frac{1}{2\sigma_0^2}\right)<\\
&<\ln \left(\frac{\epsilon_1^2}{\epsilon_0\epsilon_2}\frac{\sigma_1^{2D}}{(2\pi)^{D/2}\sigma_2^{D}}\frac{1}{(2\sigma_1^2-\sigma_2^2)^{D/2}} \right) \label{eq:separation}
\end{split}
\end{equation}
This relation can be further simplified by assuming that $\sigma_1^2=(\sigma_0^2+\sigma_2^2)/2$ and $\sigma_1=c\sigma_0$, where $c$ is the proportionality constant. Under such choice of parameters the right hand side of \eqref{eq:separation} becomes identically 0, so the inequality reads:
\begin{equation}
0<\ln\left(\tilde \epsilon^2 \frac{(1+c^2)^D}{(2\pi)^{D/2}c^D} \right) \label{eq:separation2}
\end{equation}
where $\tilde \epsilon=\epsilon_1/\sqrt{\epsilon_0\epsilon_2}$ is a common energy scale.  In Fig \ref{fig:separation} we have presented a region on $\tilde \epsilon$-$c$ plane where \eqref{eq:separation2} is satisfied for $D=3$. This plot can be directly compared with Figure 2 from \cite{bib:gauss_separation}, which presents the region of mixing and demixing. There is a significant qualitative agreement between both pictures, for $c>1$. This suggests that the presence of attractive tail is, in fact, the main factor behind the separation. However, in contrast to \cite{bib:gauss_separation}, our model predicts also a narrow region of possible demixing for $c\to0$ and $\tilde \epsilon>0$. In this limit, $G_2(r)$ reduces to point-like particles, so effective interaction is described by \eqref{eq:Upoint2}, rather than \eqref{eq:Ueff_gauss}. Indeed, this illustrates the claim that even point-like particles can induce effective interactions in 'soft' potentials.

\subsection{Yukawa particles}\label{sec:yukawa}
The Yukawa potential is commonly used in the context of DLVO theory \cite{bib:kubo} and it is believed to accurately describe the long range interaction between screened charged particles \cite{bib:crocker}. Since it is also tractable in terms of its Fourier Transform, it is an interesting example for our theory. From now on, we are interested in the $D=3$ systems, in which case, the Yukawa potential $Y(r)$ and its Fourier Transform read:
\begin{align}
Y(r)=\epsilon \sigma \frac{e^{-\kappa(r-\sigma)}}{r}&&\mathcal{Y}(k)=\frac{4\pi \epsilon \sigma e^{\kappa \sigma}}{k^2+\kappa^2} \label{eq:yukawa1}
\end{align}

Let us consider a system composed of Yukawa particles, where $\sigma_1,\epsilon_1,\kappa_1$ describe particle-depletant interaction $Y_1(r)$ and depletant-depletant interaction $Y_2(r)$ depends on $\sigma_2,\epsilon_2,\kappa_2$. Then, the effective interaction can be calculated analytically from \eqref{eq:Ueff}, namely:
\begin{equation}
\begin{split}
&U_{eff}(\Delta R)=-\frac{1}{(2\pi)^3}\int_{\tilde \Omega} d\textbf{k} e^{\imath \bf{k \Delta R}}\frac{|\mathcal{Y}_1(k)|^2}{\mathcal{Y}_2(k)}=\\
&-\frac{\epsilon_1^2 \sigma_1^2}{\epsilon_2\sigma_2} e^{-\kappa_1(\Delta R-2\sigma_1)-\kappa_2\sigma_2}\left( \frac{1}{\Delta R}-\frac{\kappa_1^2-\kappa_2^2}{2\kappa_1}\right) \label{eq:Ueff_yukawa}
\end{split}
\end{equation}

\begin{figure}
\includegraphics[width=0.95\linewidth]{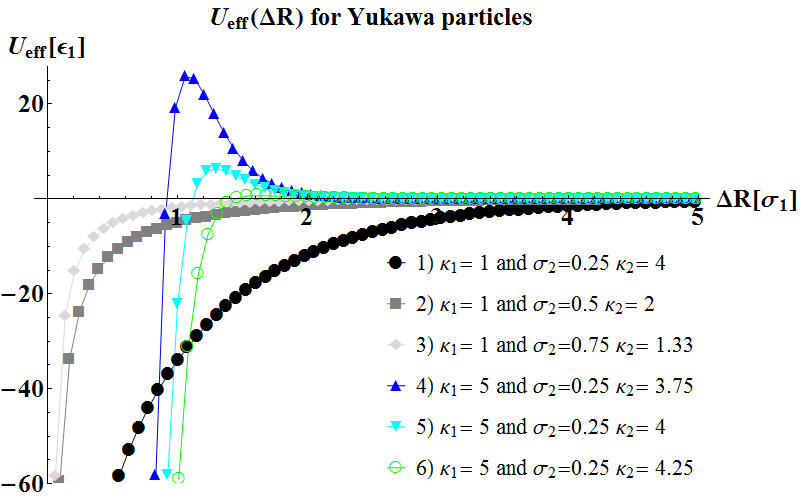}
\caption{Effective potential for binary mixture of Yukawa particles, according to formula \eqref{eq:Ueff_yukawa}, for which  $\epsilon_1=\epsilon_2>0$ and $\sigma_2$, $\kappa_i$ are given in units  $[\sigma_1]$ and $[\sigma_1^{-1}]$, respectively. Curves 1-3: growing depletion attraction for decreasing size of repellent, $\kappa_i=\sigma_i^{-1}$ to match the size of particle. Curves 4-6: for decreased particle-depletant interaction a $\kappa_2$-dependent energy barrier appears.  \label{fig:yukawa}}
\end{figure}

A graphical representation of \eqref{eq:Ueff_yukawa} for various parameters is shown in Fig.\ref{fig:yukawa}. In general, for $\epsilon_2>0$ the particular profiles of effective interaction are strongly dependent on parameters and can vary from purely attractive to strongly repulsive. When the range of interaction is of the order of particle radius ($\kappa_i\simeq \sigma_i^{-1}$), the effective interaction is attractive (curves 1-3 in Fig. \ref{fig:yukawa}) and its range increases with the downturn in the depletant radius. In fact, this range is surprisingly long, namely for $\sigma_2/\sigma_1=0.25$ the interaction is significant over a range of $5\sigma_1$ (curve 1, Fig.\ref{fig:yukawa}). This is in stark contrast with Asakura-Oosawa model for HS of radii $\sigma_1$ and $\sigma_2$, where the interaction would cease over a range of $\sigma_1+\sigma_2$ \cite{bib:yodh}. Another interesting characteristic of $U_{eff}$ for Yukawa particles appears when the range of particle-depletant interaction is decreased, in which case a repulsive barrier emerges. This barrier grows as the range of depletant-depletant interaction increases (curves 4-6, Fig.\ref{fig:yukawa}). Apparently, possible energetic advantages of lower $Y_1(r)$ cannot dominate the depletant-depletant repulsion. Finally, if we assume $\epsilon_2<0$ the global sign of $U_{eff}(\Delta R)$ is inversed, leading to repulsion-through-attraction effects.

Summarizing, this relatively simple model indicates possible self-organization of Yukawa particles, although analytical calculations analogous to Gaussian particles cannot be easily completed here. Nevertheless, phase separation in binary Yukawa systems has been encountered in simulations \cite{bib:yukawa_separation} and also in the context of plasma research, e. g. \cite{bib:plasma1, bib:plasma2}. 

\subsection{HS-like particles with Yukawa interaction tail}\label{sec:yukawa_hs}
Yukawa potential suffers from the lack of repulsive core independent from the interaction tail, so realistic description of colloid particles requires more complicated potential. In \cite{bib:louise} Louis et al. have simulated binary system consisting of HS particles with Yukawa interaction tails, both as depletant and colloid particles. For repulsive tails, \cite{bib:louise} reports the effective attraction in the system, while attractive tails induce 'repulsion through attraction'. Within our framework we are able to qualitatively reproduce these two effects with analytical formula.

We propose to model both hard core and interaction tail of a single particle with two Yukawa potentials, namely:
\begin{equation}
Y_i^{HS}(r)=\frac{c_i\sigma_i}{r}e^{-\lambda_i(r-\sigma_i)}+\frac{t_i\sigma_i}{r}e^{-\kappa_i(r-\sigma_i)}
\end{equation} 
where index $i=1$ denotes particle-depletant interaction and $i=2$ denotes depletant-depletant interaction. For $\lambda_i\gg\kappa_i$ the first term becomes impenetrable core, while the second term can be now either repulsive or attractive, depending on $t_i$. However, in order to allow direct comparison between our results and \cite{bib:louise}, we would like to control attractive tail of $Y_i^{HS}(r)$ with the depth of its minimum $\epsilon_i$. Thus, for $\epsilon_i<0$ we have determined $t_i$ numerically, from the following equations:
\begin{equation}
\left\{
\begin{aligned}
&\left.\frac{d}{dr}Y_i^{HS}(r)\right|_{r=r_0}=0 \\
&Y_i^{HS}(r_0)=\epsilon_i
\end{aligned}\right.
\end{equation}
In case of repulsive tail we have assumed $t_i=\epsilon_i\ge0$. 

The Fourier Transform of $Y_i^{HS}(r)$ is simply a sum of two $\mathcal{Y}(k)$ for relevant parameters. Therefore, the effective interaction reads:
\begin{equation}
\begin{split}
&U_{eff}(\Delta R)=-\frac{1}{(2\pi)^3}\int_{\tilde \Omega} d\textbf{k} e^{\imath \bf{k \Delta R}}\frac{|\mathcal{Y}^{HS}_1(k)|^2}{\mathcal{Y}^{HS}_2(k)} =\\
&=-\frac{2}{\pi}\frac{\sigma_1^2}{\sigma_2}\int_{0}^{+\infty}dk \frac{k\sin \Delta R k}{\Delta R}\frac{(k^2+\kappa_2^2)(k^2+\lambda_2^2)}{(k^2+\lambda_1^2)^2(k^2+\kappa_1^2)^2}\times \\
&\times\frac{\left(c_1 e^{\sigma_1 \lambda_1}(k^2+\kappa_1^2)+t_1 e^{\kappa_1\sigma_1}(k^2+\lambda_1^2)\right)^2}{\left(c_2 e^{\sigma_2 \lambda_2}(k^2+\kappa_2^2)+t_2 e^{\kappa_2\sigma_2}(k^2+\lambda_2^2)\right) } \label{eq:HS_yukawa}
\end{split}
\end{equation}
The integrand in the above expression is an even function and the degree of polynomial expression in denominator is higher than in numerator, so this integral can be calculated analytically, thanks to the residue theorem. The poles in the upper complex half plane read: 
\begin{gather*}
k_1=\imath \lambda_1\\
k_2=\imath \kappa_1\\
k_3=\imath \sqrt{\frac{c_2 \kappa_2^2 e^{\sigma_2 \lambda_2}+t_2 \lambda_2^2 e^{\sigma_2 \kappa_2}}{c_2 e^{\lambda_2 \sigma_2}+t_2e^{\kappa_2 \sigma_2}}}
\end{gather*}
The final formula is too long to be presented here, but it can be handled with the aid of symbolic algebra software.

\begin{figure}
\includegraphics[width=0.95\linewidth]{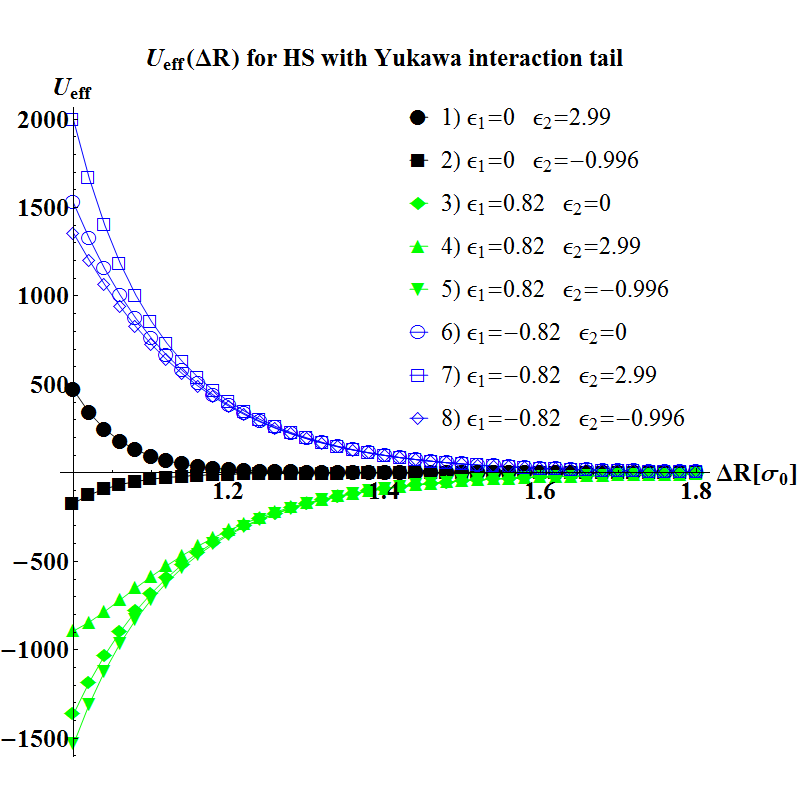}
\caption{Effective interaction in the binary mixture of hard spheres with Yukawa interaction tails, generated from formula \eqref{eq:HS_yukawa}. $\sigma_0$ is the radius of bigger particles, $\sigma_1=0.6\sigma_0$, $\sigma_2=0.2\sigma_0$, $\kappa_1=6/\sigma_0$ and $\kappa_2=15/\sigma_0$. Core parameters for all curves: $\lambda_1=10\kappa_1$, $\lambda_2=10\kappa_2$, $c_1=c_2=0.1$. Curves 1 and 2: for $\epsilon_1=0$ behavior of $U_{eff}$ depends on the sign of $\epsilon_2$. Curves 3-5: for $\epsilon_1>0$ $U_{eff}$ is attractive. Curves 5-8: for $\epsilon_1<0$ $U_{eff}$ is repulsive.\label{fig:HS_yukawa}}
\end{figure}

The selection of results generated from \eqref{eq:HS_yukawa} is presented in Fig. \ref{fig:HS_yukawa}, which is our analytical counterpart of Fig. 6 from \cite{bib:louise}. The values of parameters has been chosen according to \cite{bib:louise}, namely, $\sigma_0=1$ is the radius of bigger particle, $\sigma_1=0.6$, $\sigma_2=0.2$, $\kappa_1=6$ and $\kappa_2=15$. The values of $\epsilon_1$ and $\epsilon_2$ have been set to 0, 0.82, -0.82 and  0, 2.99, -0.996, respectively. We have also chosen the core parameters equal $\lambda_i=10\kappa_i$ and $c_1=c_2=0.1$. In case of $Y_i^{HS}(r)$ with the attractive tail these core parameters resulted in the actual radius of repulsive core within $(1\pm0.1)\sigma_i$ and the position of minimum within $(1\pm0.05)\sigma_i$.

Under such choice of parameters, our model reproduces three groups of results, found by Louis \cite{bib:louise}. For $\epsilon_1>0$ the effective interaction is attractive, regardless of $\epsilon_2$ (curves 3-5, Fig. \ref{fig:HS_yukawa}). Conversely, for $\epsilon_1<0$ the effective interaction is repulsive ('repulsion-through-attraction' effect, curves 6-8, Fig. \ref{fig:HS_yukawa}), once again regardless of $\epsilon_2$. In case of $\epsilon_1=0$ we obtain repulsion for $\epsilon_2>0$ and attraction in the opposite case (curves 1 and 2, Fig. \ref{fig:HS_yukawa})). Effective interaction for $\epsilon_1=\epsilon_2=0$ proved to be strongly dependent on the choice of $\lambda_i$ and $c_i$, but in this case $U_{eff}(\Delta R)$ is reduced to the formula \eqref{eq:Ueff_yukawa}, so we do not include this case in Fig. \ref{fig:HS_yukawa}.

 In general, our results are in qualitative agreement with \cite{bib:louise}, especially in terms of asymptotic behavior and the range of interaction. However, the curvature for $\Delta R<1.2$ is inaccurately reproduced and we find no shallow minima which seem to appear in the molecular dynamics simulations. Finally, it should be mentioned that the choice of $\lambda_i$ and $c_i$ have a remarkable influence on the exact shape of effective potential and the values of these parameters cannot be derived from the first principles. 

\section{Final remarks}
In this paper we have proposed occupation number functional as a tool to describe binary colloidal systems. We have derived a model of effective interactions alternative to Asakura-Oosawa approach, density functional theory and closure relations. In Section \ref{sec:applications} we have shown that with the aid of our formalism we are able to reproduce analytically basic characteristics of systems with Gaussian or Yukawa interactions. This supports our claim that $U_{eff}(\textbf{R}_i-\textbf{R}_j)$ can be the dominant source of effective interactions. The framework we propose is essentially different from standard tools in the field and while it is currently far less developed and not as accurate, it provides a more direct insight into how the effective interactions in the colloids arise from microscopic potentials. We have provided the discussion of assumptions and approximations which determine the limits of applicability for our theory. Further development of the occupation number functional approach might include reproducing thermodynamics of binary systems or relating this model to spatio-temporal correlations in noise in a Langevin-like description.

\acknowledgments{M. Majka acknowledges the support from the M.Smoluchowski KRAKOW SCIENTIFIC CONSORTIUM, in the framework of the KNOW scholarship.}

\end{document}